\documentclass[a4paper,fleqn,usenatbib,useAMS]{mnras}

\usepackage{mathptmx}

\usepackage[T1]{fontenc}
\usepackage{ae,aecompl}

\usepackage{graphicx}	
\usepackage{amsmath}	
\usepackage{amssymb}	

\newcommand{\logg}{\log\,g}
\newcommand{\teff}{T_{\rm eff}}
\newcommand{\bz}{\langle B_z \rangle}
\newcommand{\bzrms}{\bz_{\rm rms}}
\newcommand{\bzrmso}{\bz_{{\rm rms},1}}
\newcommand{\bzrmst}{\bz_{{\rm rms},2}}

\newcommand{\kms}{km\,s$^{-1}$}

\title[The anomalous atmospheric structure of HD\,166473]{
The anomalous atmospheric structure of the strongly magnetic Ap star HD\,166473}

\author[S. P. J\"arvinen et al.]{
S.~P.~J\"arvinen$^{1}$\thanks{E-mail: sjarvinen@aip.de},
S.~Hubrig$^{1}$,
G.~Mathys$^{2}$,
V.~Khalack$^{3}$,
I.~Ilyin$^{1}$,
\newauthor{
and
H.~Adigozalzade$^{1}$}
\\
$^{1}$Leibniz-Institut f\"ur Astrophysik Potsdam (AIP), An der Sternwarte~16, 14482~Potsdam, Germany\\
$^{2}$European Southern Observatory, Alonso de Cordova 3107, Vitacura, Santiago, Chile\\
$^{3}$D\'epartement de Physique et d'Astronomie, Universit\'e de Moncton, Moncton, NB, Canada E1A 3E9
}

\date{Accepted XXX. Received YYY; in original form ZZZ}

\pubyear{2020}

\begin{document}
\label{firstpage}
\pagerange{\pageref{firstpage}--\pageref{lastpage}}
\maketitle

\begin{abstract}
High resolution spectropolarimetric observations of the strongly magnetic,
super-slowly rotating rapidly oscillating Ap star HD\,166473 are used to
investigate the implications of the presence of a variable strong magnetic
field on the vertical and surface horizontal distribution of various chemical
elements. The analysis of the calculated LSD Stokes $I$ and $V$ profiles
confirms the previously reported detection of non-uniform horizontal surface
distribution of several chemical elements. To test the vertical abundance
stratification of iron peak and rare earth elements, magnetic field
measurements were carried out using spectral lines of these elements
belonging to neutral and ionised stages. We find clear indication of the
existence of a relation between the magnetic field strength and its
orientation and vertical element stratification: magnetic field values
obtained for elements in different stages close to the magnetic equator are
rather similar whereas the dispersion in field strengths is remarkably large
in the regions close to magnetic field poles. At the phases of negative and
positive extrema the mean longitudinal field strength determined from the
analysis of the REE lines is usually stronger than when using Fe and Cr. The
strongest mean longitudinal magnetic field, up to $-$4160$\pm$226\,G, is
detected using the \ion{La}{ii} line list at the negative extremum, followed
by the measurements using the \ion{Pr}{iii} lines with
$\bz$=$-$3740$\pm$343\,G and the \ion{Ce}{ii} lines with
$\bz$=$-$3372$\pm$247\,G. The strongest mean longitudinal magnetic field of
positive polarity, up to 3584$\pm$354\,G is detected using the \ion{Pr}{iii}
lines, followed by the measurement $\bz$=2517$\pm$249\,G using the
\ion{Ce}{ii} lines.
\end{abstract}




\section{Introduction}
\label{sec:intro}

The discovery of the presence  of resolved magnetically split lines in the 
slowly rotating, cool Ap star HD\,166473 (CoD\,$-37$~12303) was reported by 
\citet{Mathys1997}.  
The values of the mean magnetic field modulus that they measured range from 
6.4 to 8.6\,kG. 
\citet{MatHub1997} 
had previously obtained for HD\,166473 three measurements of the mean 
longitudinal field, all of the order of $-2$\,kG.
\citet{kurtzmar} 
discovered rapid oscillations in this star, whose fundamental parameters, 
$\teff=7700$\,K and $\logg=4.2$, were determined by 
\citet{Gelbmann}. 
\citet{kurtzmar} reported the existence of at least three pulsation modes 
with periods between 8.8 and 9.1\,min and rather low photometric amplitudes 
(less than 0.5\,mmag). 

Most recently, 
\citet[][hereafter MKL2020]{Mathys2020}
presented the first accurate determination of the rotation period of 
HD\,166473, $P_{\mathrm{rot}}=3836\pm30$\,d, from the analysis of 56 
measurements of the mean magnetic field modulus $\langle B \rangle$ and 21 
determinations of the mean longitudinal magnetic field $\bz$. The latter were 
based on high- and medium-resolution spectra acquired between 1992 and 2019 
at various observatories and with various instrumental configurations, among 
them high-resolution ESPaDOnS and HARPS\-pol polarimetric spectra. The value 
of the period derived by MKL2020, which is primarily based on the analysis of 
very precise measurements of the mean magnetic field modulus spanning 2.6 
rotation cycles, is strongly constrained by the availability between phases 
0.05 and 0.25 of several observations obtained at similar phases during three 
different rotation cycles. In particular, the alternative value of the period, 
$P_{\mathrm{rot}}=3893$\,d, that was proposed by 
\citet{Bychkov2020} 
is extrapolated since it is based on observations spanning slightly less than 
a rotation cycle. It is definitely too long, as the $\langle B\rangle$ 
variation curve obtained using it to plot the whole data set of MKL2020 shows 
systematic phase shifts from one cycle to the next.

HD\,166473, whose mean magnetic field modulus varies between 6 and 11\,kG,
possesses the third strongest field among roAp stars: stronger fields have 
been previously detected only in HD\,154708
\citep[24.5\,kG,][]{Hubrig2005, kurtz2006}
and HD\,92499 
\citep[8.2\,kG,][]{Elkin2010}. 
The variation curves of $\langle B \rangle$ and $\bz$ are well approximated 
by cosine waves. The $\bz$ variation curve shows sign reversals, implying 
that both magnetic poles of the star come alternately into sight over the 
rotation cycle. However, the phases of the extrema of the two field moments
are shifted with respect to each other by a small but formally significant 
amount. The difference in the values of $\langle B\rangle$ between the phases 
of the extrema of $\bz$, and the large ratio between the values of the 
$\langle B\rangle$ extrema further indicates that the structure of the 
magnetic field of HD\,166473 departs significantly from a centred dipole 
(MKL2020).

Evidence that singly and doubly ionised Pr and Nd lie in a thin layer high in 
the atmosphere of HD\,166473 was presented for the first time by 
\citet{Gelbmann}. 
Furthermore, 
\citet{cowley2001} 
identified a strong core-wing anomaly in the H Balmer lines: that is, the 
broad wings of these lines end abruptly in narrow cores. This indicates an 
extremely abnormal temperature-depth structure of the atmosphere. 
\citet{kurtz2003}  
took advantage of this anomalous atmospheric structure to study the vertical 
resolution of the pulsation modes into standing waves in the atmosphere and 
overlying running waves in the upper atmosphere. They also showed that the 
core of the H$\alpha$ line is formed deeper than the Pr and Nd lines, that 
stronger Pr and Nd lines form higher up in the atmosphere than the weaker 
ones, and that Fe lines form below the level of H$\alpha$ core formation, 
probably near radial pulsation nodes for all three modes. 

A few years ago, 
\citet{Hubrig2018} 
analysed high-quality spectropolarimetric material obtained with HARPS\-pol 
for Przybylski's star (HD\,101065), a very slowly rotating star with a  
probable period of 188\,yr and $\bz$ of several hundred Gauss. They used 
longitudinal magnetic field measurements to discuss the anomalous structure 
of the atmosphere, in particular the inhomogeneous vertical distribution of 
several chemical elements. However, because Przybylski's star rotates much 
more slowly than HD\,166473, the structure of its magnetic field remains 
undefined. The availability of the high-resolution ESPaDOnS and HARPS\-pol 
spectropolarimetric observations of HD\,166473 used in the recent magnetic 
field study of MKL2020 makes it possible now to investigate for this star the 
relation between the vertical and horizontal distribution of various chemical 
elements and the magnetic field structure.

Although the simple axisymmetric model of the magnetic field of HD\,166473 
that was obtained by MKL2020 is not meant to be physically realistic, its 
availability makes this star an excellent target for a detailed study of the 
non-uniform vertical and horizontal distributions of different chemical 
elements and their relation to the magnetic field strength and orientation. 
Such a study based on magnetic field measurements using line lists 
constructed for various elements in the neutral state and in various 
ionisation stages has the potential to provide an important test of the 
competing atomic segregation processes whose combination is generally 
believed to lead to the inhomogeneous and stratified atmospheres of the 
magnetic Ap stars.

In the following two sections we describe the observational material and the 
method of analysis. The obtained results and their discussion are presented 
in Sections 4 and 5.


\section{Observations and data reduction }\label{sect:obs}

\begin{table}
\centering
\caption{
  Logbook of the observations. The columns give the telescope and instrument
  configuration, the heliocentric Julian date (HJD), rotation phase 
  $\varphi$, and the exposure time. The rotation phase is calculated using 
  the rotation period, $P_{\mathrm{rot}}=3836$\,d, and the phase origin, 
  HJD$_{0}=2448660.0$, adopted by MKL2020.
}
\label{T:obs}
\begin{tabular}{llcc}
\noalign{\smallskip}\hline \noalign{\smallskip}
Configuration & \multicolumn{1}{c}{HJD} & $\varphi$ & Exp.\ time \\
& 2\,450\,000+ & & [s] \\
\noalign{\smallskip}\hline \noalign{\smallskip}
ESO 3.6\,m + HARPS & 6148.655& 0.952 & 1800 \\
CFHT + ESPaDOnS    & 6531.773 & 0.052 & 2060 \\
                   & 6547.732 & 0.056 & 2060 \\
                   & 6813.014 & 0.125 & 2040 \\
                   & 7239.836 & 0.237 & 1014 \\
                   & 7287.712 & 0.249 & 2028 \\
                   & 8642.993 & 0.602 & 2096 \\
\noalign{\smallskip}\hline \noalign{\smallskip}
\end{tabular}
\end{table}

The observations used in this study were already described in detail by
MKL2020. The spectropolarimetric material analysed by these authors included
seven high-resolution spectropolarimetric observations obtained at different 
epochs: six were acquired with the ESPaDOnS spectropolarimeter fed by the 
Canada-France-Hawaii Telescope (CFHT), and one with the High Accuracy Radial 
velocity Planet Searcher (HARPS) in spectropolarimetric mode fed by the 
ESO 3.6-m telescope. The logbook of the observations is presented in 
Table~\ref{T:obs}. The rotation phase is calculated using the rotation 
period, $P_{\mathrm{rot}}=3836$\,d, and the phase origin, 
HJD$_{0}=2448660.0$, adopted by MKL2020. The obtained spectra sample rather 
well the rotation cycle from phase 0.95 to phase 0.60, allowing us to analyse 
magnetic field measurements covering about two thirds of the rotation cycle 
including rotation phases close to the negative and positive extrema.


\section{Magnetic field measurements}
\label{sec:analys}

Similar to the previous study of Przybylski's star
\citep{Hubrig2018},
to measure the mean longitudinal magnetic field and to increase the 
signal-to-noise ratio (S/N), we employ the least squares deconvolution 
technique
\citep[LSD;][]{Donati1997}.
This technique assumes that the lines used in the analysis have an identical 
shape and the resulting profile is scaled according to the line strength and 
sensitivity to the magnetic field. Using the Vienna Atomic Line Database 
\citep[VALD; e.g.,][]{Kupka2011,VALD3}, 
we created 16 line masks for nine elements including the iron-peak elements 
Cr and Fe and seven rare-earth elements (REE), all based on the stellar 
parameters of HD\,166473, $\teff=7700$\,K and $\logg=4.2$ 
\citep{Gelbmann}.
The line masks for Cr and Fe include lines belonging to the neutral atoms and 
the first ions. Line lists for first and second ions were created for Ce, Pr, 
Nd, Eu, and Er, whereas for La the identified lines of the second ion are too
blended or weak and the only known \ion{Sm}{iii} lines are at wavelengths 
shorter than 3850\,\AA. Those lines of the aforementioned ions that are 
located in spectral regions contaminated by telluric lines are not included 
in the line lists.

\begin{figure}
 \centering 
        \includegraphics[width=1.0\columnwidth]{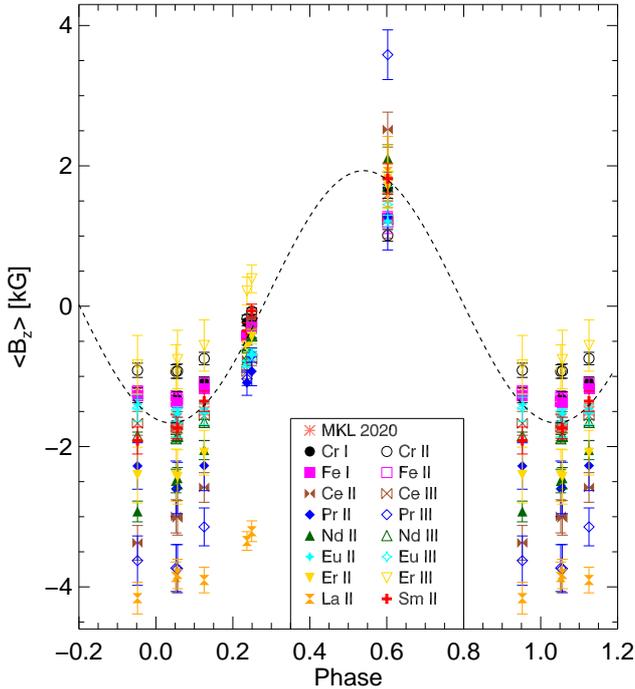}
        \caption{
         Mean longitudinal magnetic field $\bz$ values calculated for various 
         line masks at the phases listed in Table~\ref{T:obs}.
         Additionally, the $\bz$ values from MKL2020 measured using 
         exclusively \ion{Fe}{i} lines (MKL~2020) are presented by asterisks 
         and their best fit solution for the magnetic phase curve by 
         the dashed line.
        }
   \label{fig:Bphase}
\end{figure}

The mean longitudinal magnetic field is determined by computing the 
first-order moment of the LSD Stokes~$V$ profile according to
\citet[][]{Mathys1989}:

\begin{equation}
\left<B_{\mathrm z}\right> = -2.14 \times 10^{11}\frac{\int \upsilon V
  (\upsilon){\mathrm d}\upsilon }{\lambda_{0}g_{0}c\int
  [I_{c}-I(\upsilon )]{\mathrm d}\upsilon},
\end{equation}

\noindent
where $\upsilon$ is the velocity shift from the line centre, in \kms,
and $\lambda_{0}$ and $g_{0}$ are the average values of, respectively, the 
wavelengths (in nm) and the effective Land\'e  factors of all the lines used 
to compute the LSD profile.

The results of the LSD magnetic field measurements using 16 different line 
masks are presented in Table ~\ref{T:Bz} along with the number of lines in 
each mask and the average Land\'e factors $g_{0}$. In most cases the false 
alarm probability (FAP) for the field detections is less than $10^{-6}$. 
According to 
\citet{Donati1992}, 
a Zeeman profile with FAP $\leq 10^{-5}$ is considered as a definite 
detection, $10^{-5} <$ FAP $\leq 10^{-3}$ as a marginal detection, and FAP 
$> 10^{-3}$ as a non-detection. In general, all the marginal or 
non-detections found for a few elements refer to the phases where the field 
polarity is turning from negative to positive. These are the measurements for 
\ion{Er}{ii} and \ion{Eu}{iii} at $\varphi=0.237$, for which we obtain only 
marginal detections with, respectively, $\mathrm{FAP}=2.4\times10^{-5}$ and  
$\mathrm{FAP}=1.5\times10^{-5}$. For \ion{Ce}{ii}, \ion{Pr}{ii}, and 
\ion{Sm}{ii} at $\varphi=0.237$ and at $\varphi=0.249$, the FAPs are 
$> 10^{-3}$. \ion{Er}{iii} is a special case: only the last detection at
$\varphi=0.602$ is a definite one while at $\varphi=0.237$ we get a 
non-detection. All the other \ion{Er}{iii} phases give only marginal 
detections with FAPs just below $10^{-3}$. This is due to the very small 
number of clean unblended \ion{Er}{iii} lines identified in the polarimetric 
spectra.

\begin{table}
\centering
\caption{
  Longitudinal magnetic field strength measurements using the LSD technique 
  for 16 different line masks. In Column 2 we present the number of lines 
  ($N$) in the individual line mask and the average Land\'e factor. For each 
  line list, the seven rows in the Column 3 correspond to the epochs listed 
  in Table~\ref{T:obs}.
}
\label{T:Bz}
\begin{tabular}{lc r@{$\pm$}l lc r@{$\pm$}l }
\hline
 & $N$       & \multicolumn{2}{c}{$\bz$} & 
 & $N$       & \multicolumn{2}{c}{$\bz$} \\ 
 & $g_{0} $  & \multicolumn{2}{c}{[G]} &
 & $g_{0} $  & \multicolumn{2}{c}{[G]} \\
\hline
\ion{Cr}{i}   & 18   & $-$1269 & 104 & 
\ion{Cr}{ii}  & 19   & $-$913  & 102 \\ 
              & 1.27 & $-$1304 & 95  &
              & 1.25 & $-$931  & 96  \\
              &      & $-$1329 & 111 &
              &      & $-$925  & 101 \\
              &      & $-$1083 & 75  &
              &      & $-$745  & 88  \\
              &      & $-$230  & 49  &
              &      & $-$176  & 49  \\
              &      & $-$181  & 56  &
              &      & $-$70   & 44  \\
              &      & 1684    & 146 &
              &      & 1009    & 84  \\
\noalign{\smallskip}
\ion{Fe}{i}   & 120  & $-$1282 & 53 &
\ion{Fe}{ii}  & 24   & $-$1213 & 77 \\
              & 1.23 & $-$1362 & 66 &
              & 1.20 & $-$1299 & 92 \\
              &      & $-$1370 & 64 &
              &      & $-$1327 & 96 \\
              &      & $-$1173 & 50 &
              &      & $-$1100 & 81 \\
              &      & $-$417  & 31 &
              &      & $-$414  & 41 \\
              &      & $-$300  & 27 &
              &      & $-$225  & 36 \\
              &      & 1256    & 75 &
              &      & 1193    & 157 \\
\noalign{\smallskip}
\ion{Ce}{ii}  & 14   & $-$3372 & 247 &
\ion{Ce}{iii} & 7    & $-$1669 & 170 \\
              & 1.01 & $-$2995 & 236 &
              & 1.26 & $-$1720 & 193 \\
              &      & $-$3016 & 246 &
              &      & $-$1697 & 195 \\
              &      & $-$2583 & 211 &
              &      & $-$1560 & 166 \\
              &      & $-$764  & 164 &
              &      & $-$931  & 101 \\
              &      & $-$168  & 117 &
              &      & $-$750  & 95 \\
              &      & 2517    & 249 &
              &      & 1235    & 267 \\
\noalign{\smallskip}
\ion{Pr}{ii}  & 7    & $-$2277 & 331 &
\ion{Pr}{iii} & 5    & $-$3626 & 352 \\
              & 1.03 & $-$2603 & 397 &
              & 0.97 & $-$3732 & 334 \\
              &      & $-$2591 & 366 &
              &      & $-$3740 & 343 \\
              &      & $-$2272 & 356 &
              &      & $-$3145 & 270 \\
              &      & $-$1091 & 179 &
              &      & $-$1036 & 94 \\
              &      & $-$928  & 205 &
              &      & $-$696  & 103 \\
              &      & 1270    & 468 &
              &      & 3584    & 354 \\
\noalign{\smallskip}
\ion{Nd}{ii}  & 20   & $-$2925 & 148 &
\ion{Nd}{iii} & 8    & $-$1853 & 59 \\
              & 1.07 & $-$2491 & 168 &
              & 1.36 & $-$1890 & 60 \\
              &      & $-$2451 & 152 &
              &      & $-$1856 & 59 \\
              &      & $-$2050 & 137 &
              &      & $-$1658 & 45 \\
              &      & $-$737  & 102 &
              &      & $-$588  & 34 \\
              &      & $-$412  & 87 &
              &      & $-$417  & 32 \\
              &      & 2108    & 196 &
              &      & 1702    & 105 \\
\noalign{\smallskip}
\ion{Eu}{ii}  & 7    & $-$1399 & 67 &
\ion{Eu}{iii} & 4    & $-$1449 & 193 \\
              & 1.62 & $-$1568 & 65 &
              & 1.15 & $-$1676 & 184 \\
              &      & $-$1531 & 58 &
              &      & $-$1662 & 206 \\
              &      & $-$1467 & 50 &
              &      & $-$1515 & 170 \\
              &      & $-$827  & 32 &
              &      & $-$778  & 99 \\
              &      & $-$684  & 26 &
              &      & $-$687  & 82 \\
              &      & 1213    & 74 &
              &      & 1449    & 256 \\
\noalign{\smallskip}
\ion{Er}{ii}  & 10   & $-$2406  & 397 &
\ion{Er}{iii} & 3    & $-$834   & 418 \\
              & 1.11 & $-$2408  & 372 &
              & 1.31 & $-$954   & 405 \\
              &      & $-$2433  & 388 &
              &      & $-$760   & 417 \\
              &      & $-$2088  & 319 &
              &      & $-$560   & 366 \\
              &      & $-$587   & 91 &
              &      & 218      & 197 \\
              &      & $-$434   & 105 &
              &      & 391      & 198 \\
              &      & 1710     & 318 &
              &      & 1855     & 446 \\
\hline
\ion{La}{ii}  & 15   & $-$4160 & 226 &
\ion{Sm}{ii}  & 12   & $-$1914 & 192 \\
              & 1.09 & $-$3860 & 215 &
              & 1.14 & $-$1715 & 167 \\
              &      & $-$3816 & 210 &
              &      & $-$1736 & 155 \\
              &      & $-$3902 & 183 &
              &      & $-$1354 & 146 \\
              &      & $-$3346 & 133 &
              &      & $-$350  & 112 \\
              &      & $-$3207 & 148 &
              &      & $-$66   & 96  \\
              &      & 1913    & 506 &
              &      & 1817    & 214 \\
\hline
\end{tabular}
\end{table}

\begin{figure}
 \centering 
        \includegraphics[width=1.0\columnwidth]{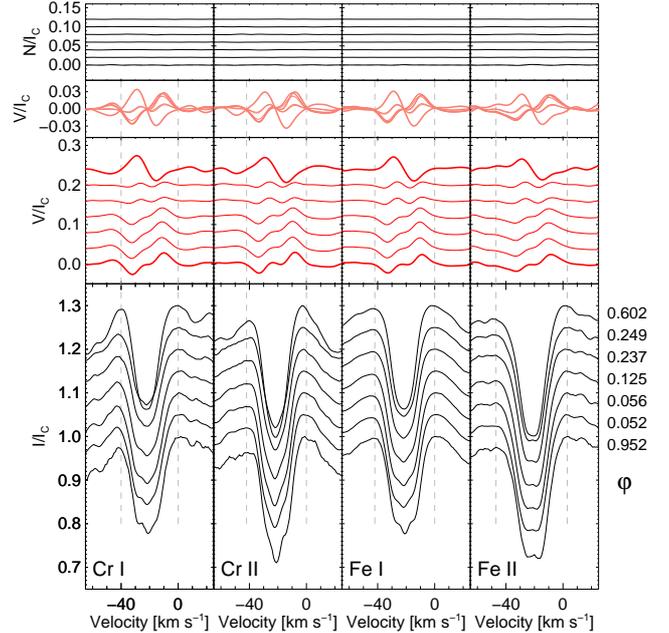}
        \caption{
         LSD Stokes $I$ (bottom), Stokes $V$ (middles), and diagnostic null 
	 ($N$) profiles (top) obtained for HD\,166473 on seven nights. The LSD 
	 profiles were calculated using \ion{Cr}{i}, \ion{Cr}{ii}, \ion{Fe}{i}, 
	 and \ion{Fe}{ii} lines. The first Stokes $V$ observation obtained 
         close to the negative field extremum and the last observation 
         obtained close to the positive field extremum are highlighted by 
         thicker red lines.
        }
   \label{fig:IVNiron}
\end{figure}

\begin{figure*}
 \centering 
        \includegraphics[width=1.0\textwidth]{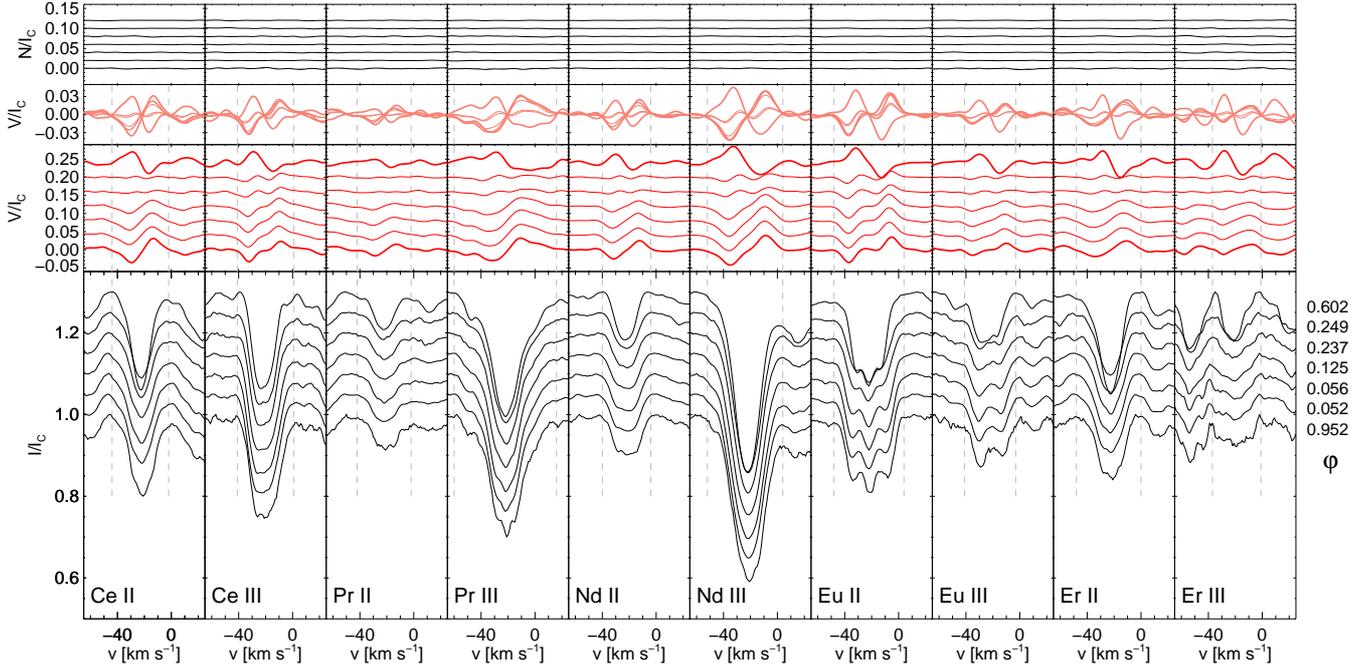}
        \caption{
          As Fig.~\ref{fig:IVNiron}, but for REEs with different 
          ionisation stages: \ion{Ce}{ii}, \ion{Ce}{iii}, \ion{Pr}{ii}, 
          \ion{Pr}{iii}, \ion{Nd}{ii}, \ion{Nd}{iii}, \ion{Eu}{ii}, 
          \ion{Eu}{iii}, \ion{Er}{ii}, and \ion{Er}{iii} lines.
          }
   \label{fig:IVNree}
\end{figure*}

\begin{figure}
 \centering 
        \includegraphics[width=1.0\columnwidth]{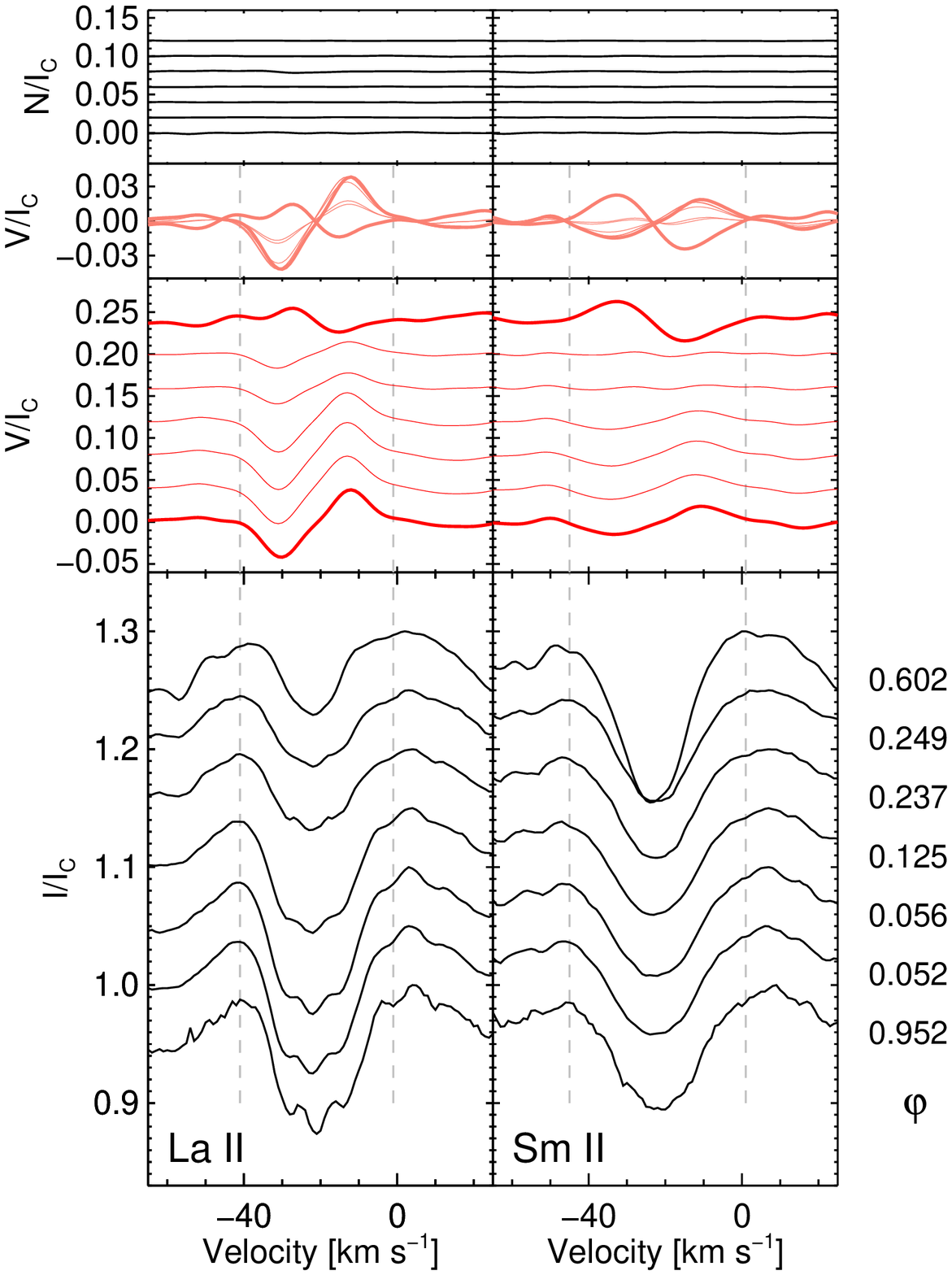}
        \caption{
         As Fig.~\ref{fig:IVNiron}, but using \ion{La}{ii}
	and \ion{Sm}{ii} line lists.
        }
   \label{fig:IVNsree}
\end{figure}

In Fig.~\ref{fig:Bphase} we display the mean longitudinal magnetic field values 
listed in Table ~\ref{T:Bz}, phased as in Table~\ref{T:obs}. In this figure 
we also show the field measurements of MKL2020, which are based exclusively 
on 20 \ion{Fe}{i} lines. The dashed curve is the best sinusoidal 
least-squares fit solution for the phase curve defined by magnetic field 
measurements presented in MKL2020. For fitting parameters and more details we 
refer to MKL2020. We note that, in general, the absolute values of the 
longitudinal field that are determined using all REE line masks are 
significantly larger close to the phases of the two extrema than the absolute 
values that are obtained using lines belonging to the Fe and Cr line masks. 
The largest absolute values of the mean longitudinal magnetic field at the 
rotation phases corresponding to the negative extremum, up to 
$-$4160$\pm$226\,G, are derived using the \ion{La}{ii} line list, followed by 
the measurements using the \ion{Pr}{iii} line list with 
$\bz$=$-$3740$\pm$343\,G and the \ion{Ce}{ii} line list with 
$\bz$=$-$3372$\pm$247\,G. The largest positive value of the mean longitudinal 
magnetic field, up to 3584$\pm$354\,G, is derived using the \ion{Pr}{iii} 
line list, followed by the measurement using the \ion{Ce}{ii} line list, with 
$\bz$=2517$\pm$249\,G.

\begin{figure}
 \centering 
        \includegraphics[width=1.0\columnwidth]{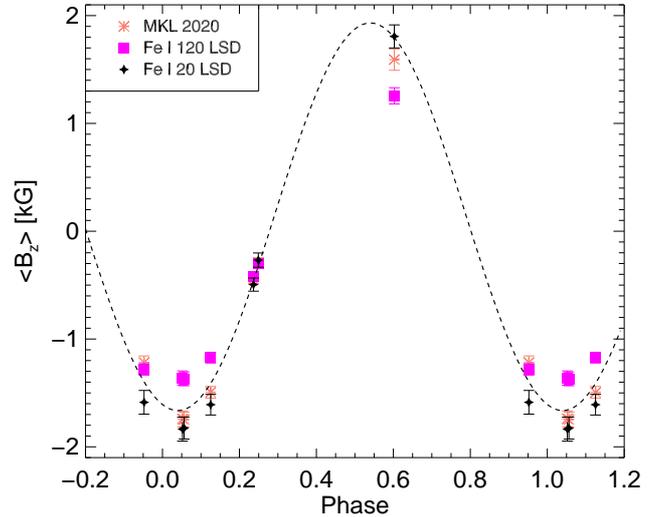}
        \caption{
         Comparison of the mean longitudinal magnetic field values
         obtained by MKL2020 by application of the moment technique
         with the LSD determinations based on the list of 20
         \ion{Fe}{i} lines of this reference and on the full list of
         120 \ion{Fe}{i} lines of this analysis.
	 The fit as in Fig.~\ref{fig:Bphase}.
        }
   \label{fig:FeIBz_e}
\end{figure}

The values of $\bz$ that are derived from the analysis of the \ion{Fe}{i} 
line list differ significantly from the values obtained by MKL2020, at least 
close to the phases of the extrema of the mean longitudinal magnetic field. 
At these phases, differences of the order of 350\,G are found between the 
results of the two analyses, which represents $\sim$20--30\% of the actual 
field values.

To gain insight into the origin of these discrepancies, we determined $\bz$ 
by application of the LSD technique to the \ion{Fe}{i} line list that MKL2020 
analysed with the moment technique. As can be seen in Fig.~\ref{fig:FeIBz_e}, 
when applied to the same line list, both methods give measurements that are 
mostly consistent within their uncertainties. The small residual differences 
can almost certainly be attributed to the different assumptions underlying 
these methods. In particular, the interpretation of the measurements made 
with the moment technique rests on the weak-line approximation. This 
approximation is robust, in that even when its conditions are not strictly 
met, the resulting measurement errors tend to remain moderate. The weak-field 
approximation on which LSD relies is, however, more fragile, as departures 
from its validity tend to entail rapidly growing measurement errors. The 
magnetic field of HD\,166473 is too strong for the weak-field approximation 
to be strictly valid. Thus, the small differences between the $\bz$ values 
derived by application to the same line list of the moment technique, on the 
one hand, and of LSD, on the other, almost certainly just reflect the 
respective limitations of the approximations on which they are based.

\begin{figure}
 \centering 
        \includegraphics[width=1.0\columnwidth]{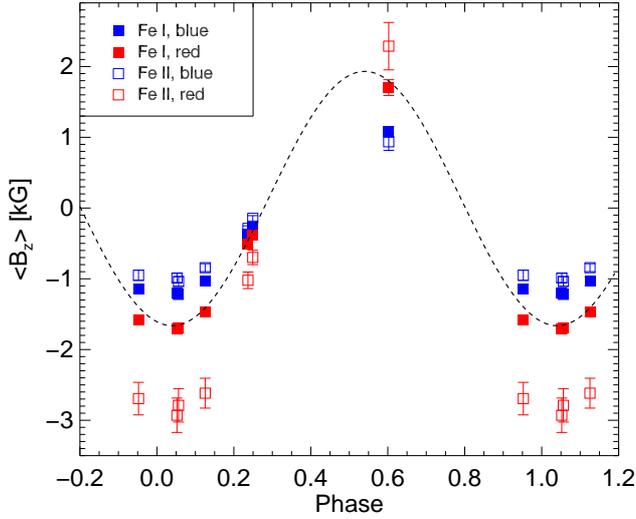}
        \caption{
         Comparison of the mean longitudinal magnetic field values obtained 
         by application of the LSD technique to subsets of lines of 
         \ion{Fe}{i} (filled squares) and \ion{Fe}{ii} (open squares) with 
         wavelengths shorter (blue symbols) and longer (red symbols) than 
         5440\,\AA.
         The fit as in Fig.~\ref{fig:Bphase}.
        }
   \label{fig:Febluered}
\end{figure}

By contrast, the discrepancies between the results of the LSD analysis of two 
sets of \ion{Fe}{i} lines are very significant. They must be attributed to 
the different characteristics of these two line sets. The list used in the 
rest of the present study comprises 120 \ion{Fe}{i} lines, while there are 
only 20 lines of this ion in the list of MKL2020. The two lists partly 
overlap, but that of MGL2020 is restricted to the wavelength interval 
5400--6800\,\AA, while the full 120-line list spans the range 
4400--6700\,\AA. As is well known, the continuum opacity in the visible changes
significantly with wavelength, so that at the red end of the spectrum, one 
can observe lines that form in deeper photospheric layers than at the blue 
end. The resulting impact on the mean longitudinal magnetic field 
measurements is illustrated in Fig.~\ref{fig:Febluered}. It shows the results 
of an experiment in which the LSD technique was applied to determine the mean 
longitudinal magnetic field of HD\,166473 from two subsets, one blue and one 
red, of the line lists for \ion{Fe}{i} and \ion{Fe}{ii}. Among the ions 
studied here, \ion{Fe}{i} and \ion{Fe}{ii} have the longest line lists. The
dividing wavelength between the two subsets was set at 5440\,\AA. There are 
77 \ion{Fe}{i} lines and 15 \ion{Fe}{ii} lines blueward of this wavelength, 
and 43 \ion{Fe}{i} lines and 9 \ion{Fe}{ii} lines redward. One can see in 
Fig.~\ref{fig:Febluered} that the absolute values of $\bz$ that are derived 
from the blue line lists are systematically smaller than those obtained from 
the red line lists. This trend is consistent with the expected behaviour of a 
dipole-like magnetic field, which should be more intense deeper in the 
stellar atmosphere than higher up, and with the fact that one can observe 
lines formed deeper in the photosphere at longer wavelengths.

From consideration of the similarity of the $\bz$ values obtained by 
application of the LSD and moment techniques to the set of (red) \ion{Fe}{i} 
lines of MKL2020 restricted to the wavelength interval 5400--6800\,\AA, and 
the significant differences between the $\bz$ values derived by applying the 
LSD technique to a blue and a red subset of \ion{Fe}{i} lines, the 
discrepancies between the mean longitudinal magnetic field measurements of 
the present study and those of MKL2020 should be ascribed primarily to the 
different characteristics of the analysed line lists, especially with respect 
to their sampling of the wavelength range.

It should also be noted that the differences between the $\bz$ values 
determined by using red and blue subsets of \ion{Fe}{ii} lines are 
considerably larger than these differences for \ion{Fe}{i}. This is probably 
related to the fact that the excitation energies of the lower levels of the 
observed \ion{Fe}{ii} transitions span a much larger range ($\sim$10\,eV) 
than those of the \ion{Fe}{i} lines ($\sim$4.5\,eV), so that the lines of the
former are formed over a wider range of photospheric depths.

The LSD Stokes $I$, $V$, and diagnostic null ($N$) profiles for different 
ions of different elements are shown in 
Figs.~\ref{fig:IVNiron}--\ref{fig:IVNsree}. Null spectra are usually obtained 
by combining the different polarizations in such a way that polarization 
cancels out. In all studied cases they are essentially flat. At phases close 
to the negative extremum, the lines of certain elements show a partly 
resolved magnetically split structure in the Stokes $I$ LSD profiles; this 
structure appears different for different ions. The \ion{Fe}{i} lines usually 
show a triplet structure while the \ion{Fe}{ii} lines are doublets. A triplet 
structure is also apparent in the LSD Stokes $I$ line profiles of 
\ion{Ce}{ii}, \ion{Pr}{ii}, \ion{Pr}{iii}, \ion{Nd}{iii}, \ion{Eu}{ii}, 
\ion{Er}{ii}, \ion{Er}{iii}, and \ion{La}{ii}, whereas the profiles of 
\ion{Ce}{iii} and \ion{Eu}{iii} show a doublet structure. An interesting 
feature of the \ion{Eu}{iii} LSD magnetically resolved doublet profiles is 
the stronger intensity of the blue components.

Because they often arise from transitions between atomic levels of high 
angular momentum $J$, the REE lines tend to have very broad anomalous Zeeman 
patterns. However, their effective Land\'e factors are in average smaller and 
span a narrower range of values than those of the lines of Fe and Cr. 
Statistically, the studied REE transitions display all the types of anomalous 
Zeeman patterns 
\citep[see Fig.~1 of][]{Mathys1989}, 
with no particular preference overall. However, on an individual basis, the 
short line lists of certain ions may show a predominant occurrence of a 
certain pattern type. For instance, 6 of the 8 lines of \ion{Nd}{iii} have a 
Zeeman pattern of Type~2, whereas this type is not represented among the 6
\ion{Eu}{ii} lines, and 5 of the 7 lines of \ion{Pr}{ii} have a Type~3 
pattern
\citep[for more details, see][]{Mathys1989}. 
Furthermore, among the lines of the same type of a given ion, some show a 
wide spread of the individual components within each of the 
$\sigma_-$, $\pi$ and $\sigma_+$ sets of components, while for others, these 
sets are compact. Among the 112 REE lines that have been analysed, only 4 
have a normal triplet pattern.

The observed LSD line profiles shown in Figs.~\ref{fig:IVNree} and 
\ref{fig:IVNsree} reflect the diversity of the Zeeman patterns and strengths 
of the individual lines from which they are computed. This is a complex 
combination that, in general, does not lend itself to simple intuitive 
interpretation. This complexity is further compounded by the effect of the 
hyperfine structure that affects several of the REE at various degrees. This 
effect may in particular be responsible for departures from symmetry of the 
Stokes $I$ profiles of some spectral lines, which form in a regime of 
Paschen-Back effect rather than Zeeman effect
\citep{Landolfi2001,Hubrig2002,Khalack2012}.

\section{Link between abnormal atmospheric composition and magnetic field 
strength and orientation}
\label{sec:abn}

The calculated LSD profiles and the analysis of longitudinal magnetic field 
measurements over the rotation cycle indicate an extremely abnormal chemical 
composition of the atmosphere of HD\,166473. In the following, we discuss in 
more detail the inhomogeneous horizontal and vertical element distribution at 
different rotation phases in relation with the different magnetic field 
strength and orientation. 

\subsection{Non-uniform horizontal element distribution}
\label{sec:hor}

\citet{Mathys2017} 
noted that the equivalent widths of the \ion{Fe}{i} lines in HD\,166473 are 
definitely variable. These line intensity variations were further 
characterised by MKL2020, who also reported and studied line-intensity 
variations for \ion{Nd}{iii}. In both cases, they attributed those variations 
to non-uniform horizontal surface distribution of the chemical elements. 
Magnetic Ap stars are known to frequently display horizontal abundance 
inhomogeneities across their surfaces. Some elements, such as REE, are 
usually observed to concentrate close to the magnetic poles, whereas other 
elements prefer the regions close to the magnetic equator. Accordingly, the 
spectral lines of elements that have different distributions across the 
stellar surface sample the magnetic field topology in different manners
\citep[e.g.,][]{kochukhov2004}.

\begin{figure*}
 \centering 
        \includegraphics[width=0.7\textwidth]{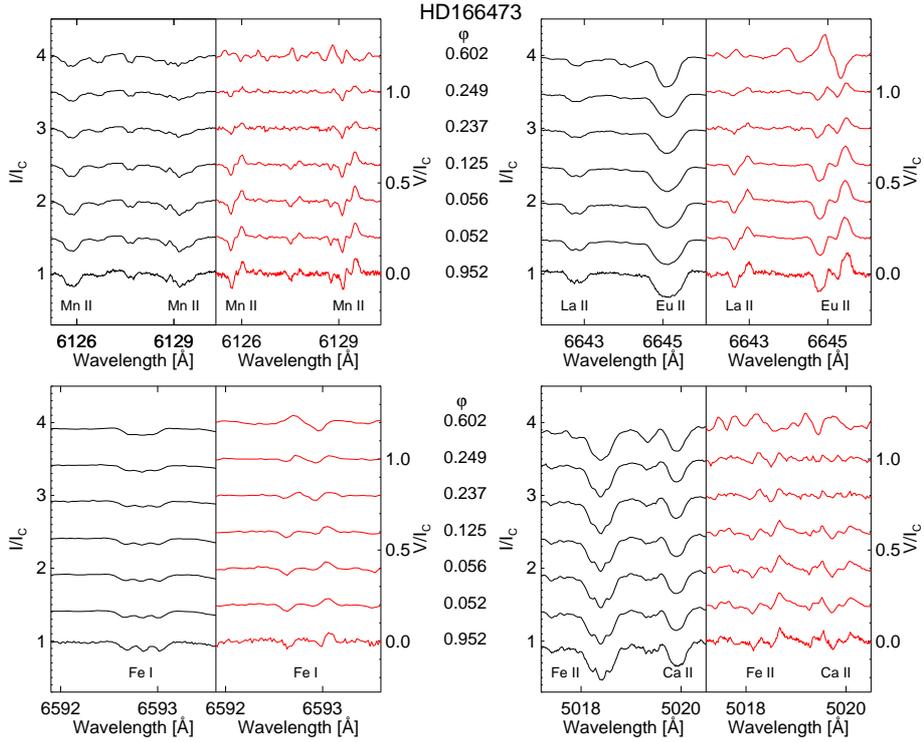}
        \caption{
          Variability of the Stokes $I$ (left) and Stokes $V$ (right) line 
          profiles belonging to a few elements in different ionisation stages 
          over the rotational cycle. 
          The top left panel shows two \ion{Mn}{ii}
          lines, the top right panel \ion{La}{ii} and \ion{Eu}{ii} lines, 
          bottom left panel \ion{Fe}{i} line, and bottom right panel 
          \ion{Fe}{ii} and \ion{Ca}{ii} lines.
                  }
   \label{fig:elemmos}
\end{figure*}

In Fig.~\ref{fig:elemmos}, we present a number of Stokes $I$ profiles of 
individual lines belonging to different elements and the corresponding Stokes 
$V$ profiles, which show a complex structure. As illustrated in 
Figs.~\ref{fig:IVNiron} and \ref{fig:IVNree}, a similarly complex structure 
with central features is also observed at the rotational phases close to the 
negative field extremum in the LSD Stokes $V$ profiles obtained for the 
\ion{Cr}{i}, \ion{Cr}{ii}, \ion{Fe}{i}, \ion{Fe}{ii}, \ion{Ce}{iii}, 
\ion{Nd}{iii}, \ion{Eu}{ii}, and \ion{Eu}{iii} line lists.

Figures~5 and 6 of MKL2020 illustrate how the presence of magnetic fields of 
opposite polarities and different strengths in the line-forming regions on 
different parts of the visible stellar hemisphere can produce structured 
Stokes V profiles. Following these authors, we attribute the complex 
structure of the Stokes $V$ profiles (both individual and LSD) of various 
ions to the presence of magnetic fields of mixed polarities over the 
hemisphere of HD~166473 that is facing us around the phase of the negative 
extremum of $\bz$. By contrast, close to phase 0.6, the Stokes $V$ line 
profiles are mostly S-shaped. Hence we expect the polarity of the magnetic 
field to be predominantly positive over most of the half of the stellar 
surface that is visible around the phase of the positive extremum of $\bz$. 

\begin{figure}
 \centering 
        \includegraphics[width=1.0\columnwidth]{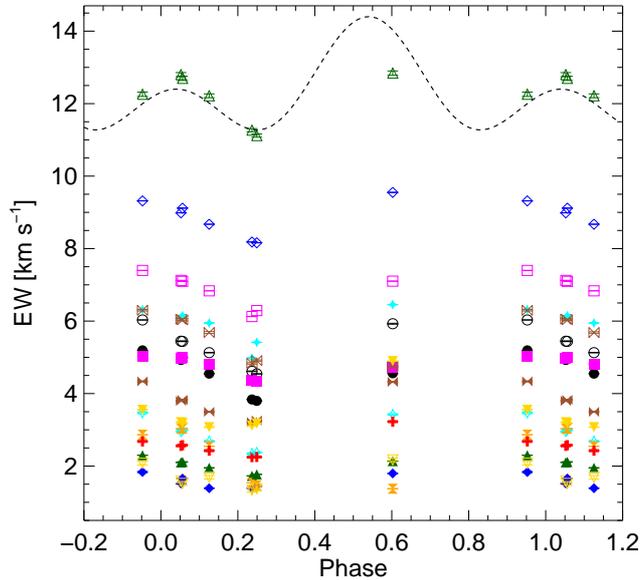}
        \caption{
	The equivalent widths (EWs) measured from LSD $I$ profiles of the 
	different elements with different ionisation stages against
        the phases from Table~\ref{T:obs}.
	The symbols are as in Fig.~\ref{fig:Bphase}. The error bars are 
        usually of the symbol size as indicated for the measurements 
        obtained using \ion{Nd}{iii} lines. The dashed line is a cosine wave 
        and its first harmonic fit obtained for \ion{Nd}{iii} 
        $\lambda 6145$\,\AA\ EWs by MKL2020 based on 56 data points.
        }
   \label{fig:ewphase}
\end{figure}

In Fig.~\ref{fig:ewphase} we present the equivalent widths (EWs) measured 
from the LSD $I$ profiles obtained for different line masks, phased as
in  Table~\ref{T:obs}. Figures~7 and 8 of MKL2020 showed the EW variability 
of the \ion{Nd}{iii} $\lambda 6145$\,\AA\ and \ion{Fe}{ii/i} 
$\lambda 6147.7$\,\AA\ lines over the rotation cycle. Two maxima and two 
minima were detected in the variation curves, with the maxima occurring close 
to the phases of extrema of the magnetic field moments $\bz$ and 
$\langle B \rangle$. Thus, both lines become stronger in the vicinity of 
magnetic poles. The primary maximum of the EW of the \ion{Nd}{iii} line 
occurs close to the positive extremum of $\bz$, while that of the Fe line 
coincides with the negative $\bz$ extremum. Our measurements of the EW of the 
LSD profiles for the line lists of the ions considered in this study, which 
are shown in Fig.~\ref{fig:ewphase}, suggest that the horizontal distribution 
of all of them is, to some extent, non-uniform. However, the phase sampling 
of our spectra is too sparse and uneven to constrain unambiguously the 
distribution of each element. While some are likely concentrated around the 
magnetic poles, similarly to Fe and Nd, we cannot definitely rule out 
different abundance patterns for some others. The fit shown in 
Fig.~\ref{fig:ewphase} is based on 56 data points of MKL2020.

\subsection{Non-uniform vertical element distribution}
\label{sec:vert}

In their abundance analysis of HD\,166473, 
\citet{Gelbmann} 
concluded that, on average, the REEs are overabundant by $+$2.8\,dex. The 
derived abundances of the second ions exceed the abundance values obtained 
from the lines of the first ions by $+$1.2\,dex for Pr, $+$1.5\,dex for Nd, 
and $+$0.7\,dex for Er. At that time, the REE anomaly had only been 
discovered in $\gamma$\,Equ 
\citep{gammaequ}
and it was found later in other roAp stars 
\citep[e.g.,][]{REEanomaly}.
The REE anomaly that is observed in HD\,166473 cannot be explained by a wrong 
absolute scale of the oscillator strengths; hyperfine structure and isotopic 
shifts cannot be responsible either for the differences between the 
abundances of the second and third REE ions 
\citep[e.g.,][]{Gelbmann, REEanomaly}.
Instead, the anomaly is usually assigned to the vertical stratification of 
the abundances of the REEs. According to 
\citet{Michaud},
in the hydrodynamically stable atmospheres of Ap and Bp stars, atomic 
diffusion can be responsible for horizontal and vertical stratification of 
abundances. Magnetic fields observed in these stars could further stabilize the 
upper atmospheric layers and significantly amplify atomic diffusion, causing 
the existence of overabundance patches in certain areas of the stellar 
atmosphere and vertical abundance  stratification
\citep{stift2016, alestift2017, alestift2019}.
In some cases, the light and iron-peak elements are concentrated in the 
deeper atmospheric layers, whereas the REEs accumulate in the upper 
atmospheric layers
\citep[e.g.,][]{cowley2000, ryabchikova2004}.

As HD\,166473 has been observed polarimetrically at seven different rotation 
phases sampling different regions of the stellar surface with different 
magnetic field strength and orientation, it appears to be an excellent target 
to investigate the implications of the magnetic field strength and 
orientation on the vertical abundance stratification of various elements. In 
particular, the available observations sample the parts of the stellar surface
around the negative and positive magnetic poles and around the magnetic 
equator. This permits a study of the abundance stratification in regions with 
very different inclinations of the magnetic field lines.

\begin{figure}
 \centering 
        \includegraphics[width=1.0\columnwidth]{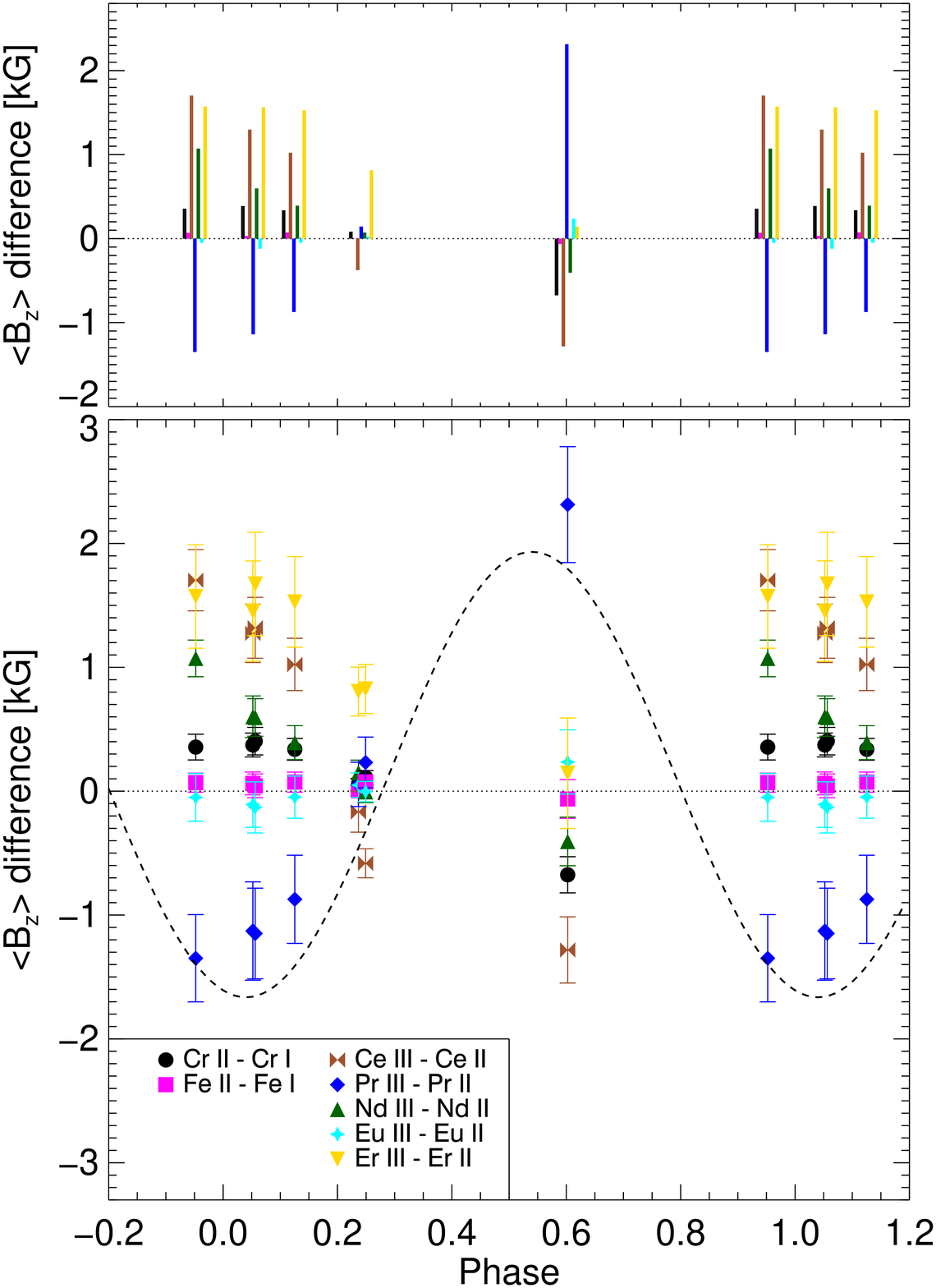}
        \caption{
	  \emph{Bottom panel:} The differences in the longitudinal magnetic 
          field measurements between the first ionization stage and the 
	  neutral stage for Cr and Fe and the second and the first ionization 
          stages for REEs. The dashed line is as in 
          Fig.~\ref{fig:Bphase}.
	\emph{Top panel:} Histogram presentation of the differences. The 
          colours of the bars are as the colours of the different symbols in 
          the lower panel and the order of the bars for each phase follows 
          the order given in the legend of the lower panel (from Cr to Er). 
          For phases where we have two almost simultaneous measurements 
          (around 0.05 and 0.24) the average value is used for clarity.
        }
   \label{fig:ionphase}
\end{figure}

To test the abundance stratification of various elements in the atmospheres in 
the strongly magnetic star HD\,166473, we analysed the differences of 
magnetic field measurements using line masks constructed for two different 
ions of various elements. The results of this experiment are shown in
Fig.~\ref{fig:ionphase}. For normal A-type star atmospheres, we expect the 
lines of elements in the higher ionisation stage to be formed lower in the 
atmosphere than those of  elements in the lower ionisation stage. Under the 
assumption that this is also the case in HD\,166473 and that the structure of 
its magnetic field is predominantly dipolar, we would expect to measure a 
stronger magnetic field in the lines belonging to the elements in the higher 
ionisation stage. We do not actually know the magnetic geometry of this
star, as the model adopted by MKL2020 was not meant to be physically 
realistic, but the arguments presented by these authors suggest that it is 
reasonable to use a dipole-like depth dependence of the photospheric magnetic 
field strength and orientation for the present analysis.

The absolute values of the mean longitudinal magnetic field determined from 
consideration of the lines of neutral Cr are systematically larger than those 
derived from analysis of the lines of the first ion. The marginal $\bz$ value 
differences between \ion{Fe}{i} and \ion{Fe}{ii} that appear in 
Fig.~\ref{fig:ionphase} may at first sight seem inconsistent with the 
behaviour illustrated in Fig.~\ref{fig:Febluered} for the measurements 
performed using separately the blue or the red subsets of lines. That the 
discrepancies between the mean longitudinal magnetic field values derived 
from the two ions with each of these subsets, which are of opposite sign, 
cancel out when the whole line list for each ion is analysed at once, must be 
considered as coincidental and partly due to the different sizes of the 
various line lists. Since 
\citet{Gelbmann} 
reported a slight overabundance of Cr and an essentially solar abundance of 
Fe, with no significant hint of abundance stratification of these elements, 
the ion-to-ion $\bz$ differences for them must primarily reflect the 
distribution of the formation depths of the diagnostic lines, as resulting 
from the excitation potential of their lower levels and from their strength.

Longitudinal field measurements using lists of lines of the first and second 
ions of REE indicate that the various elements have very different 
stratifications. While measurements using the line lists belonging to the Ce 
and Nd first ions show a significantly stronger longitudinal field close to 
the phases of the magnetic extrema, Pr shows the opposite behaviour at the 
same phases, with much stronger $\bz$ values derived from the second ion line 
list. Furthermore, the difference between the field values measured using the 
lists of the first and second ions of Pr is surprisingly large, reaching about 
1.3\,kG close to the negative extremum of the longitudinal field and 2.3\,kG 
near its positive extremum. Almost identical $\bz$ values within the 
measurement uncertainties were obtained for the first and second ions of Eu. 
Measurements using the line lists for the first and second Er ions mostly 
follow the trend detected for Ce and Nd, except in the vicinity of the 
positive magnetic pole, where the $\bz$ values are obtained from the analysis 
of the lines of either ion do not significantly differ from each other. 

The LSD Stokes $I$ and Stokes $V$ profiles of the remaining lanthanide 
elements, La and Sm, are presented in Fig.~\ref{fig:IVNsree}. For these
elements, the only blend-free lines that could be identified in the spectra 
belong to the first ionisation stage. The magnetic field of the order of 
$-$4.2\,kG measured in the vicinity of the negative magnetic pole using the 
line list for \ion{La}{ii} lines is the strongest of all the values obtained in
this analysis,  followed by $\bz=-3.7$\,kG values determined from the
analysis of \ion{Pr}{iii} lines.

\section{Discussion}
\label{sec:disc}

We compared magnetic field measurements carried out using spectral lines 
belonging to two different ionisation stages of a number of elements. This 
analysis indicates that there exists a relation between the magnetic field 
strength and orientation, on the one hand, and vertical element 
stratification, on the other hand. Namely, at rotational phases such as 
$\varphi$=0.237 and 0.249, at which HD\,166473 is observed magnetic 
equator-on, with the line of sight nearly making a right angle with the 
dipole axis, similar values of the longitudinal magnetic field values are 
obtained from consideration of the lines of the two ions of each element. By 
contrast, for observations obtained at epochs when the star is seen almost 
magnetic pole-on, very different $\bz$ values are obtained from the analysis 
of lines corresponding to different ionisation stages, especially for the 
first and second ions of Ce, Pr, Nd and Er.  

\begin{table}
\centering
\caption{
Root-mean square mean longitudinal magnetic field values determined
from $\bz$ measurements at phases 0.952, 0.052, 0.056, and 0.602 from
the spectral lines of the first and second ions of four REEs. The
results obtained by application of the LSD and moment techniques are
compared (top section). The differences between them, for a given ion
(bottom section), and the
differences between the first and second ion values, with each
measurement technique (middle section), are given. The averages of the formal
uncertainties of the $\bz$ measurements for the line lists of each ion
and for each measurement technique are also included. For the
differences, $\Delta\bzrms=\bzrmso-\bzrmst$, the values of
$\sqrt{\sigma_{\rm av,1}^2(\bz)+\sigma_{\rm av,2}^2(\bz)}$ are used as
estimates of the uncertainties ($\sigma(\Delta\bzrms)$).
}
\label{T:Bzrms}
\begin{tabular}{lrrrr}
  \hline
  \hline
&\multicolumn{2}{c}{Moment technique}&\multicolumn{2}{c}{LSD technique}\\
Ion&\multicolumn{1}{c}{$\bzrms$}&\multicolumn{1}{c}{$\sigma_{\rm av}(\bz)$}
&\multicolumn{1}{c}{$\bzrms$}&\multicolumn{1}{c}{$\sigma_{\rm av}(\bz)$}\\
   &\multicolumn{1}{c} {(G)}&\multicolumn{1}{c} {(G)}&\multicolumn{1}{c} {(G)}&\multicolumn{1}{c} {(G)}\\
\hline
\ion{Ce}{ii}&1945&180&2990&245\\
\ion{Pr}{ii}&1412&216&2252&391\\
\ion{Nd}{ii}&2009&105&2511&166\\
\ion{Er}{ii}&1989&250&2260&369\\
\\
  \ion{Ce}{iii}&1648&208&1593&206\\
  \ion{Pr}{iii}&2332&278&3671&346\\
  \ion{Nd}{iii}&1943&113&1827&71\\
  \ion{Er}{iii}&1484&527&1186&422\\
  \hline
  \hline
&\multicolumn{2}{c}{Moment technique}&\multicolumn{2}{c}{LSD technique}\\
Ion
  &\multicolumn{1}{c}{$\Delta\bzrms$}&\multicolumn{1}{c}{$\sigma(\Delta\bzrms)$}&\multicolumn{1}{c}{$\Delta\bzrms$}&\multicolumn{1}{c}{$\sigma(\Delta\bzrms)$}\\
  &\multicolumn{1}{c} {(G)}&\multicolumn{1}{c} {(G)}&\multicolumn{1}{c} {(G)}&\multicolumn{1}{c} {(G)}\\
\hline
  \ion{Ce}{iii}$-$\ion{Ce}{ii}&$-$297&275&$-$1398&320\\
  \ion{Pr}{iii}$-$\ion{Pr}{ii}&920&352&1419&522\\
  \ion{Nd}{iii}$-$\ion{Nd}{ii}&$-$66&154&$-$684&180\\
  \ion{Er}{iii}$-$\ion{Er}{ii}&$-$505&583&$-$1074&560\\
  \hline
  \hline
  &\multicolumn{2}{c}{First ion}&\multicolumn{2}{c}{Second
                                         ion}\\
  Element&\multicolumn{1}{c}{$\Delta\bzrms$}&\multicolumn{1}{c}{$\sigma(\Delta\bzrms)$}&\multicolumn{1}{c}{$\Delta\bzrms$}&\multicolumn{1}{c}{$\sigma(\Delta\bzrms)$}\\
  &\multicolumn{1}{c} {(G)}&\multicolumn{1}{c} {(G)}&\multicolumn{1}{c} {(G)}&\multicolumn{1}{c} {(G)}\\
  \hline
  Ce&$-$1045&304&55&293\\
  Pr&$-$840&446&$-$1339&444\\
  Nd&$-$502&196&117&134\\
  Er&$-$271&445&299&675\\
  \hline
\end{tabular}
\end{table}

The element-to-element differences in the discrepancies between the mean 
longitudinal magnetic field values obtained from the analysis of sets of 
lines belonging to two ionisation stages suggest that different elements have 
different vertical stratifications. Especially intriguing are the 
measurements using \ion{Pr}{ii} and \ion{Pr}{iii} lines, which suggest that 
vertical stratification for this element differs strongly from that of 
other elements. This is reminiscent of the results obtained by 
\citet{Hubrig2018} 
in their study of the vertical stratification of Nd and Pr in the atmosphere 
of Przybylski’s star. Magnetic field measurements for this star were based on 
lists of Pr and Nd lines belonging to first and second ions. A stronger 
longitudinal magnetic field was detected using lines belonging to the first 
ion of Nd. Within the measurement uncertainties, almost identical $\bz$ 
values were derived using \ion{Pr}{ii} and \ion{Pr}{iii} lines. This 
indicates that the vertical distribution of Pr in the photosphere of 
Przybylski's star greatly differs from that of Nd.

Different magnetic field strengths determined using lines of elements in 
different ionisation stages can in principle be explained by introducing
element abundance layers for different ions in the outer atmosphere at 
different geometrical depths. The formation of such layers can be related to 
the existence of a non-standard temperature gradient. Based on the work of 
\citet{Shulyak},
it is possible that if the strong overabundance of REEs is present in the 
atmosphere of HD\,166473, it can lead to the appearance of an inverse 
temperature gradient with a maximum temperature increase of up to several 
hundreds Kelvin in the upper layers compared to a homogeneous abundance 
model. Admittedly, any scenario to understand the observed differences in 
magnetic field strengths is much more complicated. Atomic diffusion amplified 
by a magnetic field is expected to cause vertical abundance stratification in 
a hydrodynamically stable atmosphere. This stratification affects the 
radiative transfer in the atmosphere, in which as a result, the temperature 
gradient and the flux redistribution become non-standard. This in turn 
influences the diffusion process, making it time-dependent or dynamical 
\citep[see][]{alestift2019}.
Clearly, more advanced theoretical simulations of atomic diffusion involving 
REEs are necessary to understand the observed anomalous atmospheric 
structures in magnetic Ap stars.

As mentioned in Sect.~\ref{sect:obs}, the magnetic model of MKL2020 is not 
meant to be physically realistic, and one should be cautious not to 
overinterpret its implications. That the actual field structure departs from 
a dipole is consistent with the view that a purely dipolar configuration is 
not stable 
\citep{Braithwaite2006},
and it is likely that, contrary to a simple dipole, the magnetic field of 
HD\,166473 is not force-free. Accordingly, the anomalous atmospheric 
structure of HD\,166473 can be additionally affected by the presence of 
electromagnetic Lorentz forces. The impact of electric currents on the 
equivalent widths of hydrogen lines was previously studied by 
\citet{Madej},
who also reported that electromagnetic forces can significantly modify the 
pressure and temperature stratification of the atmosphere, leading to 
noticeable changes of ionization degree of some elements.

In view of the difference between the values of the mean longitudinal 
magnetic field that are derived by application of the LSD technique to
different lists of \ion{Fe}{i} lines (see Sect.~\ref{sec:analys}), one may 
wonder to which extent the different values of $\bz$ that are obtained from 
the consideration of the lines of different REE ions are just an artefact of 
the analysis method. This concern is further strengthened by the great 
diversity of the Zeeman patterns of the REE diagnostic lines (see also 
Sect.~\ref{sec:analys}). Indeed, a fundamental assumption underlying the LSD 
technique is that ``most lines [$\ldots$] exhibit Zeeman signatures with more 
or less the same shape''  
\citep{Donati1997}. 
The limited validity of this assumption when lines with very different Zeeman 
patterns are analysed has an unknown impact on the derived values of the mean 
longitudinal magnetic field.

To assess in a more quantitative manner the potential implications of these
technical limitations for the physical conclusions that we derived from our 
$\bz$ measurements, we repeated these measurements applying the moment 
technique, using the line lists of the four REE elements for which the values 
of the mean longitudinal magnetic field show significant differences between 
the first and the second ion (Ce, Pr, Nd, and Er). We considered only the 
four phases closest to the magnetic extrema (0.952, 0.042, 0.056, and 0.602), 
at which the ion-to-ion $\bz$ differences are largest. For each ion, we 
computed the rms longitudinal field $\bzrms$ 
\citep{bohlender1993} 
for these four phases. As an estimate of the uncertainty affecting the $\bzrms$
values, we adopted the arithmetic mean of the formal uncertainties of the 
four individual measurements from which they were computed, 
$\sigma_{\rm av}(\bz)$. The results of this experiment appear in the top
section of Table~\ref{T:Bzrms}. In the middle section of this table, we give 
the $\bz$ differences between the two ions of each element, with each 
measurement technique. Their ``uncertainties'', $\sigma(\Delta\bzrms)$, are 
computed by application of the standard error propagation formula to the 
corresponding individual uncertainties of the rms longitudinal field values,
$\sigma_{\rm av}(\bz)$. Finally, in the bottom section of 
Table~\ref{T:Bzrms}, we present for each ion the differences between the 
$\bzrms$ values based on the field determinations through the moment 
technique and through the LSD technique. The ``uncertainties'' are computed 
in the same way as in the middle sector.

The differences $\Delta\bzrms$ between the values of the rms longitudinal 
field obtained from the analysis of the first and second ions of each of the 
considered elements are systematically larger with the LSD technique than 
with the moment technique. Actually, with the latter, $\Delta\bzrms$ is 
considerably larger than its estimated uncertainty $\sigma(\Delta\bzrms)$ only
for Pr. 
For the other REEs, the moment technique does not indicate the 
ion-to-ion $\bz$  differences that are apparent with the LSD technique. 
Furthermore, for several ions (e.g. \ion{Pr}{iii}), the difference between 
the $\bzrms$ values obtained with the two $\bz$ measurement techniques is as 
large or larger than the difference between the $\bzrms$ values determined 
from the analysis of the two ions, by application of either technique. This 
calls for caution in the interpretation of the ion-to-ion $\bz$ differences. 

The most clear-cut case is that of Pr, for which a stronger  mean longitudinal
magnetic field seems definitely obtained from consideration of the lines of 
the second ion than of the first ion, both through the moment technique and 
with LSD. The resulting inference about abundance stratification appears well 
founded. The cases of Ce, Nd, and Er are less conclusive given that the 
analysis based on the moment technique fails to fully confirm the differences
between the $\bz$ values derived from the consideration of the first and 
second ions of each element when using LSD. The occurrence of abundance 
stratification for these elements, as suggested by the LSD analysis, is 
plausible, but further evidence is required to establish it convincingly and 
characterise it better.

Figure~\ref{fig:IVNree} illustrates the LSD profiles for the lanthanide elements
Ce, Pr, Nd, Eu, and Er in the first and second ionisation stages. Among the 
studied lanthanides, \ion{Pr}{iii} and \ion{Nd}{iii} show by far the 
strongest lines. These lines are also much stronger than those of these 
elements in the first ionisation stage. Stronger lines with higher opacities 
form higher in the atmosphere
\citep{kurtz2003}.
According to 
\citet{Mashonkina2005}
and
\citet{Mashonkina2009},
the formation of the Pr and Nd lines can significantly deviate from the local 
thermo-dynamic equilibrium (LTE). Thus, the doubly ionised lines of these 
elements appear unusually strong due to the combined effects of vertical 
stratification and departures from LTE. As shown by 
\citet{Mashonkina2005}
and
\citet{Mashonkina2009},
the departures from LTE for the Pr and Nd lines of the first and the second 
ions are of the opposite sign, and they are significant if these elements are 
concentrated in the upper atmospheric layers. In these studies, the effect of 
the magnetic field was omitted from the statistical equilibrium calculations. 
More theoretical work is required to understand the impact of NLTE on the 
line formation of various elements in the presence of a strong magnetic field. 

To study the influence of a magnetic field on the atomic diffusion of 
different chemical elements in the hydrodynamically stable atmosphere of the 
Ap and Bp stars, 
\citet{stift2016}
have developed theoretical models of the time-dependent atomic diffusion. 
Depending on the strength and orientation of the magnetic field lines, 
time-dependent atomic diffusion can lead to accumulation of iron peak 
elements in the upper atmospheric layers even for a relatively weak magnetic 
field, of the order of 1~kG.
\citet{stift2016} 
have noted that the accumulation of iron in the outer atmosphere can lead to 
a temperature increase, a temperature plateau, or even a temperature 
inversion. Exploring the case of a non-axisymmetric magnetic field geometry 
\citep[][like in HD\,154708]{stiftet2013} 
\citet{alestift2017}
have modelled the 3D distribution of 16 chemical elements for equilibrium 
solutions and revealed rings or quasi-rings of enhanced Fe and Cr abundance 
in the upper atmospheric layers. 

Recently, \citet{alestift2019}
have demonstrated that in order to properly approximate the observational 
profile of the vertical stratification of elemental abundances in Ap and Bp 
stars using theoretical models of the time-dependent atomic diffusion, one 
needs to introduce a certain mass loss rate. The observational profiles of 
vertical abundance stratification derived for different chemical elements in 
the atmospheres of Ap and Bp stars in the framework of project VeSElkA  
\citep[Vertical Stratification of Element Abundances;][]{Khalack2015a, Khalack2015b} 
may help one to constrain the theoretical models of time-dependent atomic 
diffusion. Therefore a detailed analysis of the vertical stratification of 
elemental abundances in the magnetically controlled atmosphere of the very 
slowly rotating Ap star HD\,166473 will contribute to verifying 
observationally the impact of stellar rotation and magnetism on the time 
dependent-atomic diffusion, improving our understanding of this process.


\section*{Acknowledgements}

We thank the referee, Prof.\ J.\ Madej, for his useful comments.
SPJ is supported by the German Leibniz-Gemeinschaft, project
number P67-2018.
VK acknowledges support from the Natural Sciences and Engineering Research 
Council of Canada (NSERC).
Based on observation made with ESO Telescopes at the La Silla Paranal
Observatory under programme ID 089.D-0383(A).
Based on observations collected at the Canada-France-Hawaii Telescope (CFHT),
which is operated by the National Research Council of Canada, the Institut
National des Sciences de l'Univers of the Centre National de la Recherche
Scientifique of France, and the University of Hawaii.
This work has made use of the VALD database, operated at Uppsala University,
the Institute of Astronomy RAS in Moscow, and the University of Vienna.


\section*{Data Availability}

The ESPaDOnS data underlying this article are available in the CFHT Science 
Archive at https://www.cadc-ccda.hia-iha.nrc-cnrc.gc.ca/en/cfht/ and can be 
accessed with the object name. Similarly, the HARPS data are available in 
the ESO Science Archive Facility at http://archive.eso.org/cms.html. 







\bsp	
\label{lastpage}
\end{document}